\documentclass[%
floatfix,
prd,
showpacs,
nofootinbib,
superscriptaddress
]{revtex4}
\usepackage{amsmath,amssymb,here}
\usepackage[dvipdfmx]{graphicx,color}

\usepackage{aas_macros}
\usepackage{ulem}
\input{colordvi.tex}

\begin{document}
\title{The influence of primordial magnetic fields on the spherical
collapse model in cosmology}

\author{Yuki Shibusawa}
\affiliation{Department of Physics and Astrophysics, Nagoya University, Aichi 464-8602, Japan}
\author{Kiyotomo Ichiki}
\affiliation{Department of Physics and Astrophysics, Nagoya University, Aichi 464-8602, Japan}
\affiliation{Kobayashi-Maskawa Institute for the Origin of Particles and the Universe, Nagoya University,
Nagoya 464-8602, Japan}
\author{Kenji Kadota}
\affiliation{Department of Physics and Astrophysics, Nagoya University, Aichi 464-8602, Japan}

\begin{abstract}
Despite the ever growing observational evidence for the existence of the large scale magnetic fields, 
their origin and the evolution are not fully understood. If the magnetic fields are of primordial origin, 
they result in the generation of the secondary matter density perturbations and 
the previous studies show that such density perturbations enhance the number of dark matter halos.
We extend the conventional spherical collapse model by including the Lorentz force 
which has not been implemented in the previous analysis to study the evolution of density perturbations produced by primordial magnetic fields. 
The critical over-density $\delta_{\rm c}$ characterizing the halo mass function turns out to be a bigger value, $\delta_{\rm c}\simeq 1.78$, 
than the conventional one $\delta_{\rm c}\simeq 1.69$ for the perturbations evolved only by the gravitational force.

The difference in $\delta_{\rm c}$ between our model and the fully matter dominated cosmological model is small at a low redshift and, 
hence, only the high mass tail of the mass function is affected by the magnetic fields. At a high redshift,
on the other hand, the difference in $\delta_{\rm c}$ becomes large enough to suppress the halo abundance over a wide range of mass scales. 
The halo abundance is reduced for instance by as large a factor as $\sim10^5$ at $z=9$.

\end{abstract}
\maketitle

\section{Introduction}
Primordial magnetic fields (PMFs) has been intensively investigated in the literature as possible seeds for large scale magnetic fields 
observed in galaxies and clusters of galaxies (for a recent review, see \citep{2013A&ARv..21...62D}). 
Magnetic fields in galaxies at a high redshift \citep{2008Natur.454..302B} 
and in void regions \citep{2010Sci...328...73N,2010ApJ...722L..39A,2013ApJ...771L..42T} 
can well be the pieces of evidence that the seed fields are of primordial origin.  
A variety of mechanisms for PMF generation has been proposed, such as
inflation, phase transitions in the early universe and cosmological vector modes
\citep{Turner:1987bw,2012PhRvD..86b3512S,2012JCAP...10..034F,1983PhRvL..51.1488H,2013PhRvD..87h3007K,2005PhRvL..95l1301T,2006Sci...311..827I,2009CQGra..26m5014M,1970MNRAS.147..279H,2013PhRvD..87j4025S,2012PhRvD..85d3009I}.
Because the strength of generated PMFs varies from one model to another, the fields are often assumed to be random Gaussian 
and their power spectrum is assumed to follow a power law with the field amplitude $B_\lambda$ normalized at $\lambda=1$ Mpc scale 
and the power law index $n_B$. Their values has been constrained from cosmic microwave background anisotropies
observed by Planck to be $B_{\lambda}<3.4$ nG and $n_B<0$ \citep{2013arXiv1303.5076P}.

The thermal history of the universe in existence of PMFs in the early universe differs from that in the standard cosmological
model without PMFs \citep{2012PhR...517..141Y}. In particular, due to the Lorentz force exerted on the weakly ionized plasma 
after cosmological recombination, PMFs induce additional density perturbations independently from the standard adiabatic mode of perturbations. 
In this PMF-induced density mode, density perturbations in the baryon fluid are excited first and then those in the CDM fluid catch up 
with the baryon through gravitational interactions \citep{1978ApJ...224..337W,1996ApJ...468...28K,2006MNRAS.368..965T}. 
The effects of PMFs are more prominent on smaller scales where the entanglement of the field lines is stronger. 
Therefore, it is expected that PMFs with nano-Gauss strength can induce density fluctuations large 
enough to produce a larger number of clusters of galaxies \cite{2012MNRAS.424..927T}, 
stronger clustering power of Ly-$\alpha$ clouds \cite{2013ApJ...762...15P}, 
bigger cosmic shear \cite{2012ApJ...748...27P,2012JCAP...11..055F}, 
and even to realize early reionization of the universe \cite{2005MNRAS.356..778S}.

In this paper, we extend previous studies by taking account of the non-linear density perturbation evolutions to develop a spherical
collapse model with PMFs. The simplest spherical collapse model for a CDM universe was invented in pioneering works by
\cite{1969PThPh..42....9T,1972ApJ...176....1G}, and has been improved by 
incorporating various physical ingredients such as the spatial curvature, the cosmological constant,
modified gravities, effects of baryon perturbations, and clustering of massive neutrinos (see, e.g.,
\cite{1991MNRAS.251..128L,1998ApJ...508..483W,Naoz:2006ye,2003ApJ...597..645O,2010PhRvD..81f3005S,2006MNRAS.368..751N,2006A&A...454...27B,2009MNRAS.393L..31F,2011JCAP...10..038B}).
The spherical collapse model is a simple and powerful tool to study the non-linear dynamics of the gravitational collapse 
and has a wide range of applications. For example, the halo mass function is characterized by the peak height 
$\nu\equiv \delta_{\rm c}/\sigma(M,z)$, where $\sigma(M,z)$ is the variance of fluctuations of halo mass $M$ 
and redshift $z$, and $\delta_{\rm c}$ is the critical over-density extrapolated from the linear theory 
when the corresponding non-linear over-density region collapses which has been conventionally estimated using the spherical collapse model.
In the previous studies on the structure formation including PMFs, only the effect on $\sigma(M,z)$ from the PMFs 
has been considered and that on $\delta_{\rm c}$ has been ignored. 
Therefore one of the purposes of this paper is to estimate $\delta_{\rm c}$ in cosmologies with PMFs. 
We will show that the value of $\delta_{\rm c}$ can be far away from the canonical value of $\delta_{\rm c}\approx1.686$ 
for the redshifts of $z\gtrsim 3$, and the PMF effect on $\delta_{\rm c}$ can not be ignored in estimating 
the number of clustering halos at a high redshift.

This paper is organized as follows.
Section II introduces the evolution equation for the spherical collapse including the Lorentz force 
and section III discusses the virial equilibrium in existence of the Lorentz force. 
Section IV summarizes the matter power spectrum of the density perturbation due to PMFs. 
Section V shows our numerical results, followed by the conclusions in section VI.
Throughout this paper, we fix the cosmological parameters to those
derived from Planck results, i.e., $\Omega_{\rm m,0}h^2=0.147$, $\Omega_{\rm b,0}h^2=0.021$, $B_\lambda=1{\rm nG}$ and $n_B=-2.9$.

\section{Spherical collapse model}
We consider the spherical collapse model described by a top-hat spherical over-dense region and an uniform background matter density field. 
The over-dense region can be characterized by the physical halo radius $R_\textrm{ini}\equiv R(t_\textrm{ini})$ at the initial time $t_\textrm{ini}$ and the uniform initial matter density, 
\begin{equation}
\rho_i(t_{\rm ini})=\bar{\rho}_i(t_{\rm ini})(1+\delta_{i,{\rm ini}}),
\end{equation}
where the subscript $i$ denotes two matter components, dark matter and baryon, 
and  $\bar{\rho}_i(t)$ and $\delta_i$ are the mean matter density and the density perturbation, respectively.
For each component, the mass $M_i$ in the spherical over-dense region is conserved,
\begin{equation}
M_i=\frac{4\pi}{3}R_i^3\bar{\rho}_i(1+\delta_i)=\frac{4\pi}{3}R^3_{i,{\rm ini}}\bar{\rho}_{i,{\rm ini}}(1+\delta_{i,{\rm ini}})={\rm constant}.
\label{mass}
\end{equation}

\begin{figure}[tbp]
 \centering
 \includegraphics[width=10cm,clip]{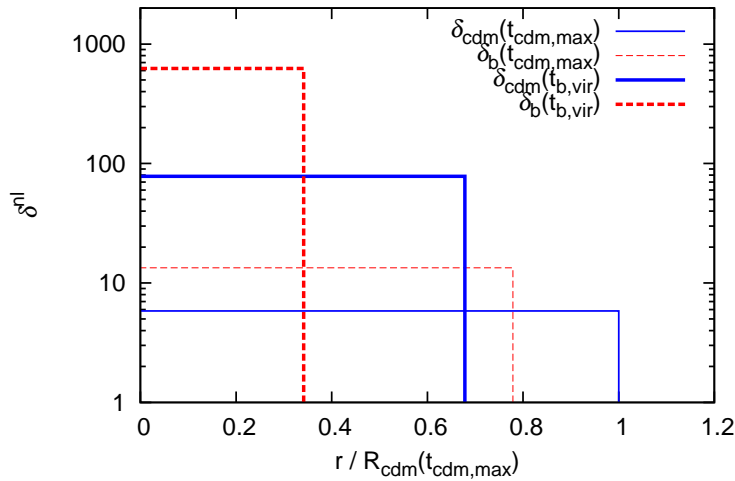}
 \caption{The spherical density contrast of our model.\: 
The vertical axis shows the density perturbations and the horizontal axis shows the radius of the spherical over-dense regions, 
normalized by the value of the dark matter over-density at the turnaround time.
The thin lines are the profiles at $t=t_{\rm cdm,max}$, when the expansion of dark matter over-dense region stops, 
and the thick ones are at t=$t_{\rm b,vir}$, when the baryon over-dense region reaches the virial equilibrium,
 and solid and dashed lines are the over-dense regions of dark matter and baryon, respectively. }
 \label{mosikizu}
\end{figure}

Assuming spherical symmetry and top-hat density distribution in the continuity, Euler and Poisson equations, 
we can obtain following non-linear differential equations of dark matter and baryon density perturbations, 
\begin{equation}
\ddot{\delta}_{\rm dm}+2H\dot{\delta}_{\rm dm}-\frac{4}{3}\frac{\dot{\delta}_{\rm dm}^2}{1+\delta_{\rm dm}}=4\pi G(1+\delta_{\rm dm})\left[\bar{\rho}_{\rm dm}\delta_{\rm dm}+\frac{R_{\rm b}^3}{R_{\rm dm}^3}\bar{\rho}_{\rm b}\delta_{\rm b}\right],
\label{dm_nonlinear}
\end{equation}
\begin{equation}
\ddot{\delta}_{\rm b}+2H\dot{\delta}_{\rm b}-\frac{4}{3}\frac{\dot{\delta}_{\rm b}^2}{1+\delta_{\rm b}}=4\pi G(1+\delta_{\rm b})\left[\bar{\rho}_{\rm dm}\delta_{\rm dm}+\bar{\rho}_{\rm b}\delta_{\rm b}\right]-3\frac{\langle F_{\rm mag}\rangle}{\bar{\rho}_{\rm b}R_{\rm b}},
\label{b_nonlinear}
\end{equation}
where the subscripts dm and b denote dark matter and baryon.
Fig.\ref{mosikizu} shows a spherical density contrast of our model.
Because the baryon density perturbations grow faster than the dark matter perturbations 
for structure formation caused by PMFs \cite{Tashiro:2009hx},
the spherical baryon over-dense region is always smaller than the dark matter over-dense region.
This effect takes the form of ${R_{\rm b}^3}/{R_{\rm dm}^3}$ in Eq.\eqref{dm_nonlinear}.
$\langle F_{\rm mag}\rangle$ is an angle averaged Lorentz force, 
and this term is added only to the equation of baryon perturbations that are subject to the influence of magnetic fields.
Some studies related to the effects of magnetic fields on the spherical collapse have been done \cite{Gopal:2010jc,Chiueh:1994}, 
and these works considered the radial Lorentz force. We also assume that the Lorentz force has only the radial component on average.

To keep the top-hat profile for the over-dense region, the Lorentz force should have the same scaling as the gravitational force.
First, we consider the spatial dependence of magnetic fields to determine the form of the Lorentz force.
The gravitational force in the matter dominated universe is written as,
\begin{equation}
F_{\rm g}=-\frac{GM\rho}{r^2}=-\frac{4}{3}\pi G\rho^2r \ \ \ (0\leq r\leq R),
\end{equation}
and this is proportional to $r$ since $\rho$ is constant inside the region of spherical over-density.
The Lorentz force reads as,
\begin{equation}
\langle F_{\rm mag}\rangle=-\frac{1}{8\pi}\frac{\partial}{\partial r}\langle B^2\rangle,
\end{equation}
where $\langle B^2\rangle$ is the dispersion of magnetic fields strength.
To march the radial scalings of the Lorentz force and gravitational force, 
we assume that $\langle B^2\rangle$ is proportional to $r^2$, i.e.,
\begin{equation}
\langle B^2\rangle=B^2(t)\left(\frac{r}{R}\right)^2 \ \ \ (0\le r\le R).
\end{equation}
Next, we focus on the time dependence of magnetic fields, $B(t)$.
Magnetic fields decay as $\propto R^{-2}$ for a comoving observer on the expanding spherical shell. 
The time dependence of magnetic fields, normalized by the initial strength of magnetic fields $B_{\rm ini}$ and the initial radius $R_{\rm ini}$, 
can then be written as,
\begin{equation}
B^2(t)=B_{\rm ini}^2\left(\frac{R_{\rm ini}}{R}\right)^4.
\end{equation}
We hence get the dispersion of the magnetic field strength and the angle averaged Lorentz force as,
\begin{eqnarray}
\langle B^2\rangle&=&B^2_{\rm ini}\left(\frac{R_{\rm ini}}{R}\right)^4\left(\frac{r}{R}\right)^2 \ \ \ (0\le r\le R),\\
\label{magnetic_strength}
\langle F_{\rm mag}\rangle&=&-\frac{1}{4\pi}B_{\rm ini}^2\left(\frac{R_{\rm ini}}{R}\right)^4\frac{r}{R^2}.
\label{Lorentz force}
\end{eqnarray}
For the evolution of radius for the outmost shell, we can consider $r=R$ in Eq.\eqref{Lorentz force}.
Since we can derive the relation 
$(R_{\rm b,ini}/R_{\rm b})^3=\bar{\rho}_{\rm b}(1+\delta_{\rm b})/\bar{\rho}_{\rm b,ini}$
from Eq.\eqref{mass}, the term that corresponds to the Lorentz force can be rewritten as,
\begin{eqnarray}
-3\frac{\langle F_{\rm mag}\rangle}{\bar{\rho}_{\rm b}R_{\rm b}}
=\frac{3}{4\pi}\frac{B_{\rm ini}^2(1+\delta_{\rm b})^2}{\bar{\rho}_{\rm b,ini}R_{\rm b,ini}^2}\left(\frac{a_{\rm ini}}{a}\right)^3.
\end{eqnarray}
We consequently arrive at the evolution equation of the baryon density perturbations with PMFs,
\begin{equation}
\ddot{\delta}_{\rm b}+2H\dot{\delta}_{\rm b}-\frac{4}{3}\frac{\dot{\delta}_{\rm b}^2}{1+\delta_{\rm b}}=4\pi G(1+\delta_{\rm b})\left[\bar{\rho}_{\rm dm}\delta_{\rm dm}+\bar{\rho}_{\rm b}\delta_{\rm b}\right]+\frac{3}{4\pi}\frac{B_{\rm ini}^2(1+\delta_{\rm b})^2}{\bar{\rho}_{\rm b,ini}R_{\rm b,ini}^2}\left(\frac{a_{\rm ini}}{a}\right)^3.
\label{b_nonlinear_3}
\end{equation}

\section{Virialization}
The over-dense region expands with the background matter density field, with a slower rate than the background expansion rate due to its self gravity.
The expansion of the over-dense region eventually stops and begins to collapse, and eventually reaches the virial equilibrium. 
The conventional virial radius for a spherical over-dense region is known to be the half of the radius at turnaround. In our model, 
the evolution of baryons is different from that of dark matter due to the Lorentz force 
and the magnetic energy should be taken into account for the baryon over-dense region. 
We hence consider the virial equilibrium for the baryon and the dark matter over-dense regions separately in the following.

We first consider the virial equilibrium for the baryon over-dense region.
We can write the virial theorem for the matter component in the spherical over-dense region of radius $R_{\rm b}$ as,
\begin{equation}
2K(R_{\rm b})+\Omega_{\rm g}(R_{\rm b})+\Omega_{\rm mag}(R_{\rm b})+\Omega_{\rm surf}(R_{\rm b})=0,
\label{virial_b}
\end{equation}
where $K$, $\Omega_{\rm g}$, $\Omega_{\rm mag}$ and $\Omega_{\rm surf}$ are 
the kinetic energy, the gravitational potential energy, the magnetic energy and  the surface term, respectively.
The radius $R_{\rm b}$ is the virial radius for baryons, $R_{\rm b,vir}$.
Both of the kinetic energies of dark matter and baryon are included in $K$,
so this term can be written  as, 
\begin{eqnarray}
K(R_{\rm b})=\frac{1}{2}\int_0^{R_{\rm b}}\rho_{\rm dm}v^2_{\rm dm}dV
+\frac{1}{2}\int_0^{R_{\rm b}}\rho_{\rm b}v^2_{\rm b}dV
=\frac{2\pi}{5}\left[\rho_{\rm dm}\left(\frac{\dot{R}_{\rm dm}}{R_{\rm dm}}\right)^2
+\rho_{\rm b}\left(\frac{\dot{R}_{\rm b}}{R_{\rm b}}\right)^2\right]R_{\rm b}^5.
\end{eqnarray}
The gravitational potential of the uniform spherical region is written as,
\begin{equation}
\Omega_{\rm g}=-\frac{3}{5}\frac{GM^2(\le R_{\rm b})}{R_{\rm b}},
\end{equation}
and the magnetic energy and the surface term can be described as,
\begin{eqnarray}
\label{mag1}
\Omega_{\rm mag}&=&\frac{3}{8\pi}\int_0^{R_{\rm b}}\langle B^2\rangle dV
=\frac{3}{10}\frac{B_{\rm ini}^2R_{\rm b,ini}^4}{R_{\rm b}},\\
\Omega_{\rm surf}&=&-\frac{1}{8\pi}\int_S\langle B^2\rangle \vec{r}\cdot d\vec{S}
=-\frac{1}{2}\frac{B_{\rm ini}^2R_{\rm b,ini}^4}{R_{\rm b}},
\label{surf1}
\end{eqnarray}
where we assume that the magnetic tension does not contribute to the magnetic energy for simplicity.
Putting all the terms together, we obtain, 
\begin{equation}
\frac{4\pi}{5}\left[\rho_{\rm dm}\left(\frac{\dot{R}_{\rm dm}}{R_{\rm dm}}\right)^2+\rho_{\rm b}\left(\frac{\dot{R}_{\rm b}}{R_{\rm b}}\right)^2\right]R_{\rm b}^5-\frac{3}{5}\frac{GM^2(\le R_{\rm b})}{R_{\rm b}}-\frac{1}{5}\frac{B_{\rm ini}^2R_{\rm b,ini}^4}{R_{\rm b}}=0.
\end{equation}
Note that the last term is the sum of Eqs.\eqref{mag1} and \eqref{surf1} representing the contributions of magnetic fields and it has the same negative sign as that of the gravitational potential term.

In the late time universe when the dark energy starts to dominate the total energy density,
one should take into account its potential energy in the virial theorem(Eq.\eqref{virial_b}).
In this paper, however, we omit the dark energy contribution in the virial theorem for simplicity 
because our main interest is in the structure formation in the early universe when the PMF effect is large.
We have checked that including the dark energy contribution to the virial condition does not change our results significantly. 
For the effect of dark energy on the virialization we refer readers to, e.g.,  
Horellou $\&$ Berge\cite{Horellou:2005qc} and Wang\cite{Wang:2005ad}.

Next, we consider the epoch after the virialization of baryons and before that of dark matter.
In the isolated system, the density and the size of the object do not change once the system reaches the virial equilibrium.
In our model, the dark matter component is still growing when the baryon reaches the virial equilibrium,
and it gets out of (or, comes into) the baryon over-dense region.
Then, the size of the baryon over-dense region changes to satisfy Eq.\eqref{virial_b}.
We can write an equation for small changes of energies in the baryon over-dense region as,
\begin{equation}
2\Delta U_{\rm b}+2\Delta K_{\rm dm}+\Delta\Omega_{\rm g}+\Delta\Omega_{\rm mag}+\Delta\Omega_{\rm surf},
\label{virial_delta}
\end{equation}
where we use the fact that the kinetic energy of baryon $K_{\rm b}$ has been converted to the internal energy $U_{\rm b}$ 
after the baryon over-dense region has virialized.
Consider the situation where dark matter within the radii $r=R_{\rm b}$ and $r=R_{\rm b}+v_{\rm dm}\Delta t$ enters the baryon over-dense region within the radius $R_{\rm b}$, and the baryon over-dense region consequently expands by $\Delta R_{\rm b}$ in a time interval $\Delta t$.
The variation of the energy included in this system is equivalent to the work done by the Lorentz force, $W$.
Then, the energy conservation implies that,
\begin{eqnarray}
\frac{1}{2}\int_{R_{\rm b}}^{R_{\rm b}+v_{\rm dm}\Delta t}\rho_{\rm dm}(t)v^2_{\rm dm}(t)dV
+\int_{R_{\rm b}}^{R_{\rm b}+v_{\rm dm}\Delta t}\rho_{\rm dm}(t)\Phi(t) dV+W
=\Delta U_{\rm b}+\Delta K_{\rm dm}+\Delta\Omega_{\rm g},
\label{energy_conserve}
\end{eqnarray}
where
\begin{eqnarray}
W&=&\int_t^{t+\Delta t} dt^\prime \int_0^{R_{\rm b}} \dot{r}\langle F_{\rm mag}\rangle dV,\\
\Delta K_{\rm dm}&=&\frac{1}{2}\int_0^{R_{\rm b}+\Delta R_{\rm b}}\rho_{\rm dm}(t+\Delta t)v^2_{\rm dm}(t+\Delta t)dV
-\frac{1}{2}\int_0^{R_{\rm b}}\rho_{\rm dm}(t)v^2_{\rm dm}(t)dV,\\
\Delta \Omega_{\rm g}&=&\int_0^{R_{\rm b}+\Delta R_{\rm b}}\rho_{\rm tot}(t+\Delta t)\Phi(t+\Delta t) dV
-\int_0^{R_{\rm b}}\rho_{\rm tot}(t)\Phi(t) dV.
\end{eqnarray}
The first two terms on the left hand side in Eq.\eqref{energy_conserve} are the kinetic and the gravitational potential energies of dark matter
which comes into the baryon over-dense region.
Substituting Eq.\eqref{virial_delta} to Eq.\eqref{energy_conserve} to eliminate $\Delta U_{\rm b}$ and $\Delta K_{\rm dm}$, we obtain,
\begin{eqnarray}
\frac{1}{2}\int_{R_{\rm b}}^{R_{\rm b}+v_{\rm dm}\Delta t}\rho_{\rm dm}(t)v^2_{\rm dm}(t)dV
+\int_{R_{\rm b}}^{R_{\rm b}+v_{\rm dm}\Delta t}\rho_{\rm dm}(t)\Phi(t) dV+W
=\frac{1}{2}\Delta\Omega_{\rm g}-\frac{1}{2}\Delta\Omega_{\rm mag}-\frac{1}{2}\Delta\Omega_{\rm surf}.
\end{eqnarray}
In our spherical collapse model, each term can be expressed as,
\begin{eqnarray}
\frac{1}{2}\int_{R_{\rm b}}^{R_{\rm b}+v_{\rm dm}\Delta t}\rho_{\rm dm}(t)v^2_{\rm dm}(t)dV
&\approx&\frac{1}{2}(4\pi R_{\rm b}^2\rho_{\rm dm}v_{\rm dm}\Delta t)v_{\rm dm}^2=\frac{1}{2}\Delta mv_{\rm dm}^2,\\
\int_{R_{\rm b}}^{R_{\rm b}+v_{\rm dm}\Delta t}\rho_{\rm dm}(t)\Phi(t) dV
&\approx&-\frac{GM(\le R_{\rm b})}{R_{\rm b}}\Delta m,\\
W&\approx&-\frac{B_{\rm ini}^2R_{\rm b,ini}^4}{R_{\rm b}^7}\Delta R_{\rm b}\int_0^{R_{\rm b}} r^4dr
=-\frac{1}{5}\frac{B_{\rm ini}^2R_{\rm b,ini}^4}{R_{\rm b}^2}\Delta R_{\rm b},\\
\Delta \Omega_{\rm g}&\approx&-\frac{6}{5}\frac{GM(\le R_{\rm b})}{R_{\rm b}}\Delta m
+\frac{3}{5}\frac{GM^2(\le R_{\rm b})}{R_{\rm b}^2}\Delta R_{\rm b},\\
\Delta\Omega_{\rm mag}&\approx&-\frac{3}{10}\frac{B_{\rm ini}^2R_{\rm b,ini}^4}{R_{\rm b}^2}\Delta R_{\rm b},\\
\Delta\Omega_{\rm surf}&\approx&\frac{1}{2}\frac{B_{\rm ini}^2R_{\rm b,ini}^4}{R_{\rm b}^2}\Delta R_{\rm b}.
\end{eqnarray}
where we approximate that $\Delta r\approx \dot{r}\Delta t$ 
and the expansion rate $H\equiv\dot{r}/r$ is constant for the baryon over-dense region, i.e.,
\begin{equation}
\frac{\Delta r}{r}=\frac{\Delta R_{\rm b}}{R_{\rm b}}= H\Delta t.
\end{equation}
Then, we can relate the radius variation to the amount of change for the mass in the baryon over-dense region as, 
\begin{equation}
\Delta R_{\rm b}=\frac{5R_{\rm b}^2}{3GM^2(\le R_{\rm b})+B_{\rm ini}^2R_{\rm b,ini}^4}
\left(v_{\rm dm}^2-\frac{4}{5}\frac{GM(\le R_{\rm b})}{R_{\rm b}}\right)\Delta m.
\label{delta_r_b}
\end{equation}
We use this equation to calculate $R_{\rm b}$ after the virial equilibrium for the baryon over-dense region
to obtain the baryon density perturbations.

Finally, we consider the virial equilibrium for the dark matter over-dense region.
The equilibrium is realized when $R_{\rm dm}$ satisfies the relation,
\begin{equation}
2U_{\rm b}(R_{\rm b})+2K_{\rm dm}(R_{\rm dm})+\Omega_{\rm g}(R_{\rm dm})
+\Omega_{\rm mag}(R_{\rm b})+\Omega_{\rm surf}(R_{\rm b})=0.
\label{virial_dm}
\end{equation}
The kinetic energy of baryon is converted to the internal energy since the baryon over-dense region has already virialized.
Substitution of Eq.\eqref{virial_b} to this equation leads to
\begin{equation}
\int_{R_{\rm b}}^{R_{\rm dm}}\rho_{\rm dm}v^2_{\rm dm}dV+\int_{R_{\rm b}}^{R_{\rm dm}}\rho_{\rm dm}\Phi dV=0.
\end{equation}
These terms can be calculate as,
\begin{eqnarray}
\int_{R_{\rm b}}^{R_{\rm dm}}\rho_{\rm dm}v^2_{\rm dm}dV&=&
\frac{4\pi}{5}\rho_{\rm dm}\left(\frac{\dot{R}_{\rm dm}}{R_{\rm dm}}\right)^2(R_{\rm dm}^5-R_{\rm b}^5),\\
\int_{R_{\rm b}}^{R_{\rm dm}}\rho_{\rm dm}\Phi dV&=&
-\frac{3}{5}\frac{GM_{\rm dm}M_{\rm tot}}{R_{\rm dm}^6}(R_{\rm dm}^5-R_{\rm b}^5).
\end{eqnarray}
Thus, we get,
\begin{equation}
\left[\frac{4\pi}{5}\rho_{\rm dm}\left(\frac{\dot{R}_{\rm dm}}{R_{\rm dm}}\right)^2-\frac{3}{5}\frac{GM_{\rm dm}M_{\rm tot}}{R_{\rm dm}^6}
\right](R_{\rm dm}^5-R_{\rm b}^5)=0.
\label{virial_dm2}
\end{equation}
Because the dark matter density perturbations grow more slowly than the baryon density perturbations,
we consider the whole system to be virialized when the dark matter over-dense region collapses
and satisfies the condition of Eq.\eqref{virial_dm2}, namely, when
the radius of dark matter over-dense region catches up with that of the
baryon's, $R_{\rm dm}=R_{\rm b}$.

\section{Density perturbation produced by primordial magnetic fields} 
After recombination, PMFs generate secondary baryon density perturbations \cite{Kim:1994zh,Mack:2001gc}.
The baryon density perturbations then induce the dark matter density perturbations through gravitational force.
In this section, we calculate the linear power spectrum of these density
perturbations that will be used to calculate the halo mass function in the following discussion.
Under the assumption that there is no correlation between PMFs and primordial density perturbations,
the power spectrum can be written as \cite{Tashiro:2009hx},
\begin{equation}
P(k,t)=P_{\rm P}(k,t)+P_{\rm M}(k,t),
\end{equation}
where the first term on the right hand side $P_{\rm P}(k,t)$ is the power spectrum of the standard adiabatic density perturbations 
and the second term $P_{\rm M}(k,t)$ is that of the density perturbations generated by PMFs. 
We can describe $P_{\rm M}(k,t)$ as, 
\begin{equation}
\label{pmk}
P_{\rm M}(k,t)=\left(\frac{\Omega_{\rm b}}{\Omega_{\rm m}}\right)^2\left(\frac{t_{\rm ini}^2}{4\pi\Omega_{\rm b}\rho_{\rm cri,0}a^3(t_{\rm ini})}\right)^2D_{\rm M}^2(t)I^2(k),
\end{equation}
where $\rho_{\rm cri,0}$ is the critical density at present time, $t_{\rm ini}$ is the initial time which is set to the recombination epoch,
$D_{\rm M}(t)$ is the growth rate and, 
\begin{equation}
I^2(k)\equiv\langle|\nabla\cdot(\nabla\times\vec{B}_0(\vec{x}))\times\vec{B}_0(\vec{x})|^2\rangle.
\label{I^2}
\end{equation}
Here $\vec{B}_0(\vec{x})$ is the comoving strength of the magnetic fields. 
Under the assumption of isotropic Gaussian statistics for primordial magnetic fields, 
the non-linear convolution of Eq.\eqref{I^2} is rewritten as\cite{Kim:1994zh},
\begin{equation}
I^2(k)=\int dk_1\int d\mu \frac{P_B(k_1)P_B(|\vec{k}-\vec{k}_1|)}{|\vec{k}-\vec{k}_1|^2}[2k^5k_1^3\mu+k^4k_1^4(1-5\mu^2)+2k^3k_1^5\mu^3],
\label{Kim et al.}
\end{equation}
where $\mu=\vec{k}\cdot\vec{k}_1/|\vec{k}||\vec{k}_1|$ and $P_B(k)\propto k^{n_B}/k_{\rm c}^{n_B+3}$ is the magnetic field spectrum
parametrized by the power law index $n_B$ and an ultraviolet cutoff scale $k_{\rm c}$.
We can analytically estimate Eq.\eqref{Kim et al.} in the limit of $k/k_{\rm c}\ll1$ as,
\begin{equation}
I^2(k)\sim\alpha\langle B_0^2\rangle^2\frac{k^{2n_B+7}}{k_{\rm c}^{2n_B+6}}+\beta\langle B_0^2\rangle^2\frac{k^4}{k_{\rm c}^3},
\end{equation}
where $\alpha$ and $\beta$ are the coefficients which depend on $n_B$, 
and $k_{\rm c}$ is written as \cite{Jedamzik:1996wp,Subramanian:1997gi},
\begin{equation}
k_{\rm c}=\left[143\left(\frac{B_\lambda}{1{\rm nG}}\right)^{-1}\left(\frac{h}{0.7}\right)^{1/2}\left(\frac{\Omega_{\rm b}h^2}{0.021}\right)^{1/2}\right]^{2/(n_B+5)}{\rm Mpc}^{-1},
\end{equation}
 in the matter dominated epoch.
 
 We introduce an important scale for the evolution of density perturbations called the magnetic Jeans scale.
 The density perturbations below this scale can not grow due to the magnetic pressure gradients.
 The magnetic Jeans scale reads as \cite{Kim:1994zh},
 \begin{equation}
k_{\rm MJ}=\left[12.5\left(\frac{B_\lambda}{1{\rm nG}}\right)^{-1}\left(\frac{\Omega_{\rm m}h^2}{0.147}\right)^{1/2}\right]
^{2/(n_B+5)}{\rm Mpc}^{-1}.
\end{equation}
 We assume that the density perturbations below this scale do not grow.

\section{Numerical Results}
\subsection{Radius of spherical over-dense region}
\begin{figure}[tp]
\begin{minipage}[m]{0.49\textwidth}
\rotatebox{0}{\includegraphics[width=1.0\textwidth]{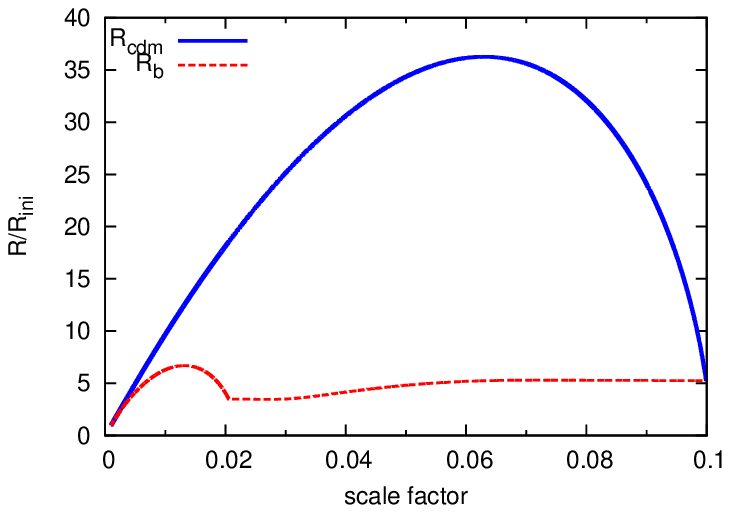}}
\end{minipage}
\begin{minipage}[m]{0.49\textwidth}
\rotatebox{0}{\includegraphics[width=1.0\textwidth]{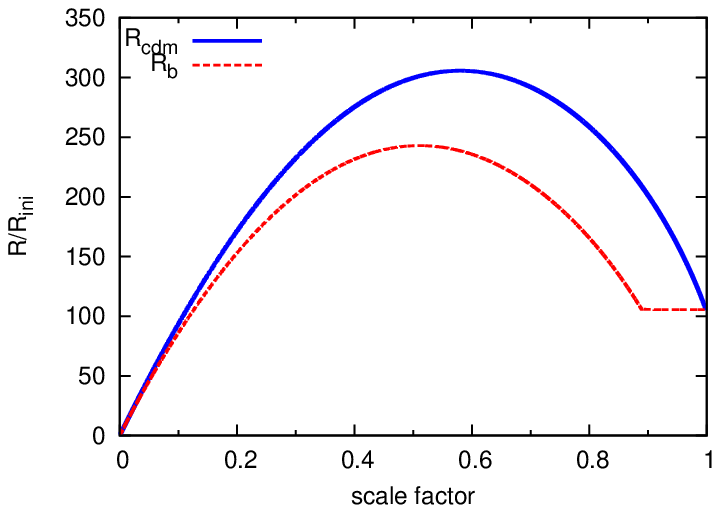}}
\end{minipage}
\caption{The radii of the over-dense regions normalized by the initial value as a function of the scale factor.\:
The left panel is for the collapse time at $a_{\rm coll}=0.1$,
and the right panel is for $a_{\rm coll}=1.0$.
The horizontal axis shows the scale factor and the vertical axis shows the radius normalized to its initial value.
The blue solid line is the radius of the dark matter over-dense region, and the red dashed line is that of the baryon over-dense region.
}
 \label{R}
\end{figure}

In Fig.\ref{R}, we show the radii of the over-dense regions, normalized by the initial value, as a function of the scale factor.
The left panel in the figure shows the case where the system collapses at a high redshift, $a_{\rm coll}=0.1$,
and the right panel shows the case at a low redshift, $a_{\rm coll}=1.0$.
We pay a particular attention to the evolution of the baryon over-dense region size after it is virialized.
After the baryon over-dense region reaches the virial equilibrium, the radius of the baryon over-dense region changes according to Eq.\eqref{delta_r_b}.
If the baryon over-dense region collapses at a low redshift, 
the dark mater over-dense region is already contracting and the dark matter is falling into the baryon over-dense region at the baryon collapse time. This leads to the gravitational potential in the baryon over-dense region, and it starts to contract. However, the enough gravitational source already exists inside the baryon over-dense region, and the baryon over-dense region contraction due to the infalling dark matter is not significant. If, on the other hand, the baryon over-dense region collapses at a high redshift, the dark matter over-dense region is still expanding and the dark matter gets out of the baryon over-dense region. In this case, the baryon over-dense region loses the gravitational potential support and results in the expansion in contrast to the contraction for the low redshift collapse.

\begin{figure}[tp]
 \centering
 \includegraphics[width=10cm,clip]{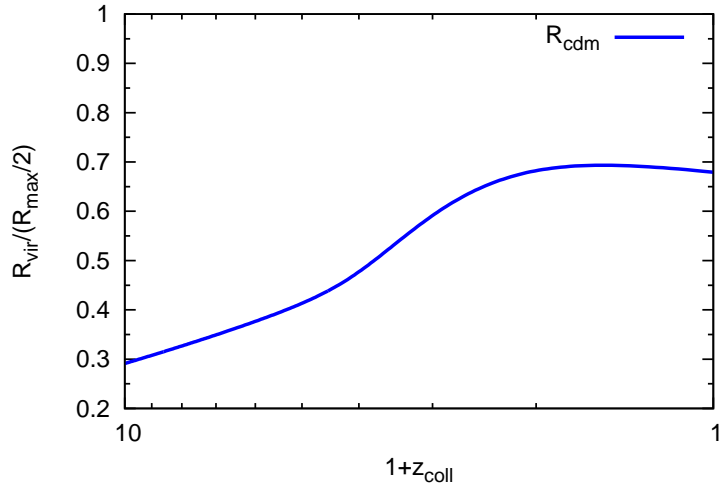}
\caption{The ratio between the virial radius in existence of PMFs and the conventional virial radius for the dark matter over-dense region.\:
The horizontal axis shows the collapse time and the vertical axis shows the ratio between the virial radius $R_{\rm vir}$, 
which we defined to be the radius at the moment of collapse in existence of PMFs, 
and the conventional virial radius defined to be the half of the radius at turnaround $R_{\rm max}/2$.}
 \label{R_vir}
\end{figure}

Fig.\ref{R_vir} depicts the ratio between the virial radius, which we defined to be the radius at the moment of collapse in existence of PMFs, and the half of the radius at turnaround, which defines the conventional virial radius, for the dark matter over-dense region. 
The dark matter over-dense region reaches the virial equilibrium when its radius catches up with that of the baryon over-dense region.
The baryon over-dense region stays smaller than the dark matter over-dense region due to PMFs,
and becomes even smaller at the higher redshift when the influence of PMFs is more powerful. The virial radius for the dark matter over-dense region hence is always smaller than $R_{\rm max}/2$.

\subsection{Critical over-density}
We calculate the critical over-density $\delta_{\rm c}$ in this subsection.
Since the critical over-density is defined as linearly evolved density perturbations at the time of virialization for the whole system, 
it is necessary to solve both the non-linear and the linear differential equations.
Therefore we linearize Eqs.\eqref{dm_nonlinear} and \eqref{b_nonlinear}
to obtain the linearized density contrast of dark matter and baryons, $\delta_{\rm dm}^{\rm lin}$ and $\delta_{\rm b}^{\rm lin}$.
Since we consider two matter components, $\delta_{\rm c}$ is written as,
\begin{equation}
\delta_{\rm c}=f_{\rm dm}\delta_{\rm dm,coll}^{\rm lin}+f_{\rm b}\delta_{\rm b,coll}^{\rm lin},
\end{equation}
where $f_{\rm dm}$ and $f_{\rm b}$ are the fractions of dark matter and baryon to the  total matter, respectively.
Similarly, we also calculate the virial over-density as,
\begin{equation}
\Delta_{\rm vir}=f_{\rm dm}\delta_{\rm dm,coll}^{\rm non-lin}+f_{\rm b}\delta_{\rm b,coll}^{\rm non-lin}.
\end{equation}
 
We consider the structure formation model caused only through the gravitational force ($\Lambda$CDM model)
to compare with our model (PMF model).
In the $\Lambda$CDM model, the initial density perturbation of dark matter at the recombination epoch is 
substantially larger than that of baryon, and it grows by self gravity of dark matter.
We assume that the virial radius in the $\Lambda$CDM model is the half of the turnaround radius.

\begin{figure}[tp]
\begin{minipage}[m]{0.49\textwidth}
\rotatebox{0}{\includegraphics[width=1.0\textwidth]{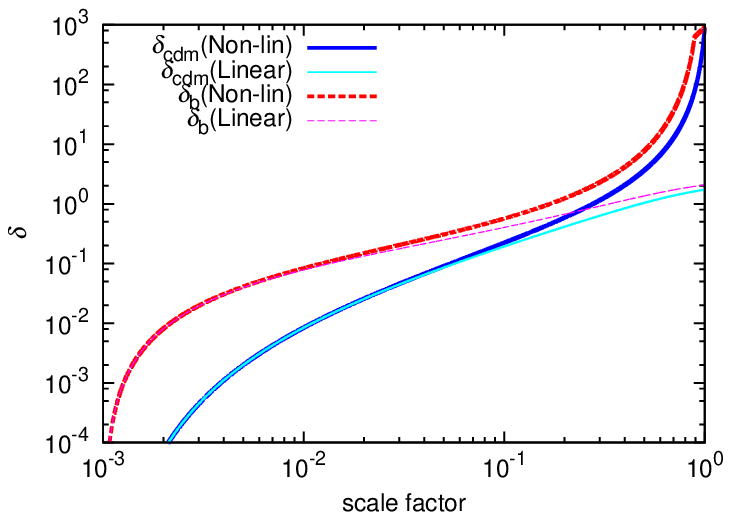}}
\end{minipage}
\begin{minipage}[m]{0.49\textwidth}
\rotatebox{0}{\includegraphics[width=1.0\textwidth]{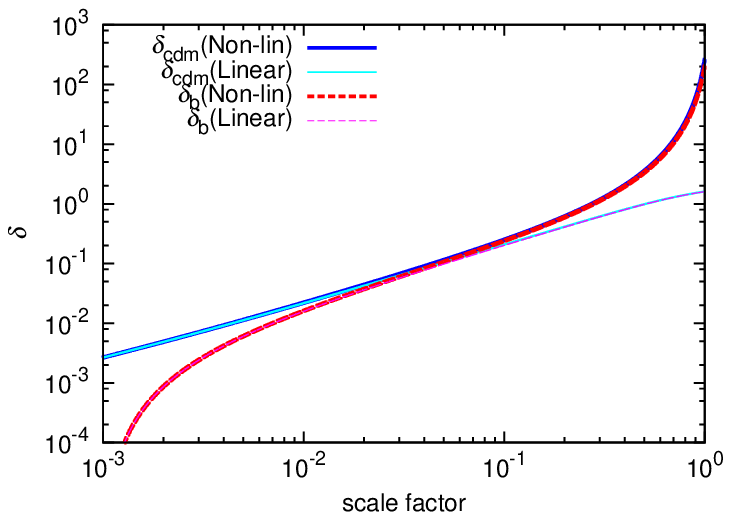}}
\end{minipage}
\caption{The time evolutions of density perturbations in the PMF model (left) and the $\Lambda$CDM model (right).\:
The thick lines are non-linear evolutions and the thin lines are linear ones,
and the solid and dashed lines are for dark matter and baryon perturbations, respectively.
} \label{delta}
\end{figure}

Fig.\ref{delta} shows the time evolution of the density perturbations where the collapse time is set at $a_{\rm coll}=1$.
The left panel in the figure shows the evolution of the density perturbations in the PMF model, 
and the right panel shows that in the $\Lambda$CDM model.
In the PMF model, both linear and non-linear evolutions of baryon density perturbations 
are enhanced by PMFs at the early stage of the evolution.
After that, the dark matter density perturbations evolve gradually by the gravity of baryons.
However, dark matter density perturbations can not catch up with baryon density perturbations 
until the baryon over-dense region reaches the equilibrium.
We can calculate the critical over-density as $\delta_{\rm c}(a_{\rm coll}=1)\simeq1.78$ in this model.
In the $\Lambda$CDM model, the dark matter density perturbations already exist at the time of recombination.
The baryon perturbations catch up with the dark matter perturbations immediately, and evolve along with them.
The critical over-density in this model becomes smaller than that in the PMF model, $\delta_{\rm c}(a_{\rm coll}=1)\simeq1.61$.
In the fully matter dominated cosmological models, $\delta_{\rm c}$ is given as $\delta_{\rm c}\simeq 1.69$,
and this is the value at the time when non-linear evolution of density perturbations goes to infinity.
In our calculation, however, we define $\delta_{\rm c}$ at the time of virialization
and non-linear density perturbations at this point are $\delta^{\rm non-lin}\sim200$.
Therefore, our result is smaller than the conventional value of 1.69. 

\begin{figure}[tp]
\begin{minipage}[m]{0.49\textwidth}
\rotatebox{0}{\includegraphics[width=1.0\textwidth]{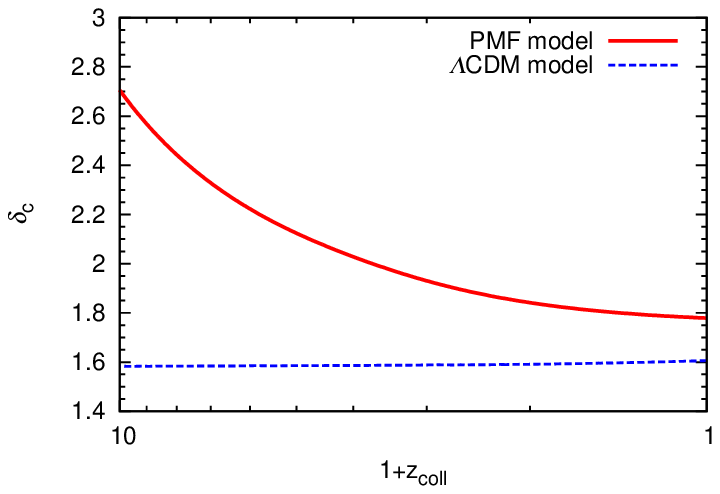}}
\end{minipage}
\begin{minipage}[m]{0.49\textwidth}
\rotatebox{0}{\includegraphics[width=1.0\textwidth]{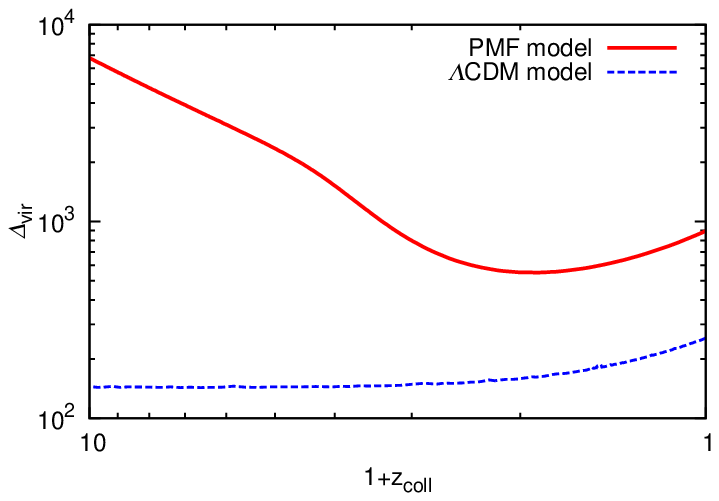}}
\end{minipage}
\caption{The critical over-densities $\delta_{\rm c}$ (left) and the virial over-densities $\Delta_{\rm vir}$ (right) at various collapse times.
The red solid line is the value in the PMF model
and the blue dashed line is that in the $\Lambda$CDM model.}
\label{a_coll}
\end{figure}

The results in Fig.\ref{a_coll} are the critical over-densities $\delta_{\rm c}$ 
and the virial over-densities $\Delta_{\rm vir}$ at various collapse times.
The left panel in the figure shows $\delta_{\rm c}$ and the right panel shows $\Delta_{\rm vir}$.
In the PMF model, $\delta_{\rm c}$ becomes larger as the system collapses earlier.
This is because the effect of magnetic fields is more significant in the linear evolution than in the non-linear stage,
and therefore linearly evolved perturbations become large at the collapse time.  
In the $\Lambda$CDM model, the change of $\delta_{\rm c}$ is small, and $\delta_{\rm c}$ becomes smaller at an earlier collapse time.
Because the evolution of the matter density perturbations is suppressed 
by radiation components at a higher redshift and by dark energy at a lower redshift,
it is shown that the value of $\delta_{\rm c}$ is reduced compared to
the canonical value $\delta_{\rm c}\simeq1.69$ \cite{Naoz:2006ye}.
In our calculation, since we consider that the collapse is completed before non-linear over-density goes to infinity 
and the effect of dark energy becomes important, the reduction of $\delta_{\rm c}$ is not significant compared to the
results in Naoz \& Barkana \cite{Naoz:2006ye}. 
The virial over-density $\Delta_{\rm vir}$ in the PMF model becomes larger than that in the $\Lambda$CDM model,
and it is about 5 times larger at $z_{\rm coll}=0$ and 60 times larger at $z_{\rm coll}=9$.
Thus, combining with the result in Fig.\ref{R_vir}, the structure formed in PMF model is denser and more compact than that in $\Lambda$CDM model.
The behavior of $\Delta_{\rm vir}$ is analogous in both models at a low redshift because $\Delta_{\rm vir}$ is mainly determined by the gravity
of dark matter and the background expansion accelerated by dark energy at a low redshift.
On the other hand, since the effect of magnetic fields becomes bigger as the collapse occurs earlier, 
$\Delta_{\rm vir}$ becomes larger at a higher redshift.

\subsection{Mass function}
The critical over-density is the threshold value for non-linear gravitational collapses,
and it is used as an input parameter for semi analytic theories of structure formation.
We use the Press-Schechter mass function\cite{Press:1974} rather than more sophisticated mass functions
\cite{oai:arXiv.org:astro-ph/9901122,oai:arXiv.org:astro-ph/0607150,oai:arXiv.org:0803.2706}.
We chose simplest mass function because our aim in this subsection is not to predict the precise number of collapsed halos,
but to roughly illustrate how large suppression of the number of halos is expected
if we use the value of $\delta_{\rm c}$ obtained in the previous section instead of the canonical value of $\delta_{\rm c}\simeq1.69$.
The mass function in the PS formalism is written as,
\begin{equation}
\frac{dn(M,z)}{dM}=\sqrt{\frac{2}{\pi}}\frac{\rho_{\rm m}}{M}\frac{\delta_{\rm c}}{\sigma^2(M,z)}\left|\frac{d\sigma(M,z)}{dM}\right|
\exp\left(-\frac{\delta_{\rm c}^2}{2\sigma^2(M,z)}\right),
\end{equation}
where $n(M,z)$ is the number density of dark matter halos which has the mass of M at redshift z,
and $\sigma(M,z)$ is the mass dispersion at mass scale $M$.
The mass dispersion is calculated from the linear matter power spectrum as,
\begin{equation}
\sigma^2(M,z)=\int dkk^2P(k,z)W(kR),
\end{equation}
where $R$ is the scale which encompasses the mass $M$ and $W(x)$ is the top-hat window function.
We use the power spectrum $P_{\rm M}(k,z)$ of the density perturbations generated by PMFs given by Eq. \eqref{pmk}.

\begin{figure}[tbp]
 \centering
 \includegraphics[width=10cm,clip]{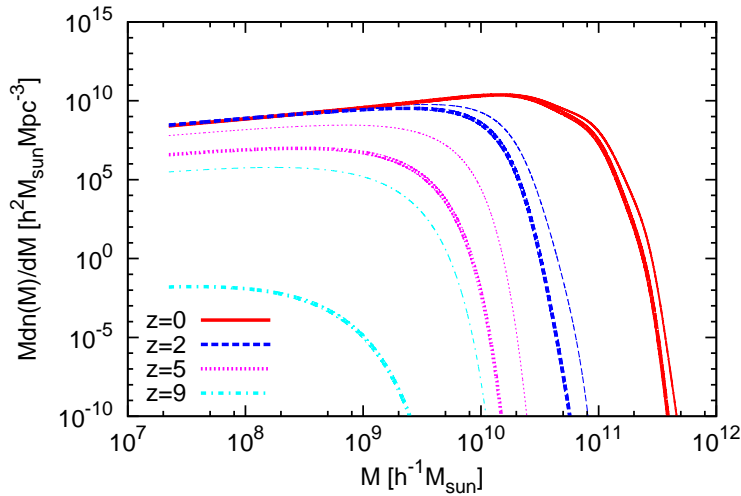}
 \caption{The mass function for the power spectrum of the density perturbations generated by PMFs.\:
The thick lines are the mass functions by using the critical over-densities $\delta_{\rm c}$ calculated in the PMF model. 
The mass functions using $\delta_{\rm c}$ values in the conventional $\Lambda$CDM model 
instead of those values calculated in the PMF model are also shown in the thin lines for comparison.
The values of $\delta_{\rm c}$ are given in Table I.}
 \label{mf}
\end{figure}

We show in Fig.\ref{mf} the PS mass function for $P_{\rm M}(k,z)$ at various redshifts.
The thick lines are the results for $\delta_{\rm c}$ calculated in the PMF model (solid line in Fig.\ref{a_coll}),
and the thin lines are the results for that calculated in the $\Lambda$CDM model (dashed line in Fig.\ref{a_coll}).
At low redshifts, the difference in $\delta_{\rm c}$ between the PMF model and the $\Lambda$CDM model is small
and it becomes appreciable only at the high mass tail of the mass function.
At high redshifts, on the other hand, the difference in $\delta_{\rm c}$ becomes large enough to 
suppress the abundance of halos over a wide range of mass scales.
The suppression becomes as large as $\sim10^5$ at $z=9$.
Thus, naive estimates of the number of halos using PS mass function with the canonical critical over-density $\delta_{\rm c}\simeq1.69$
would lead to the significant overestimates at high redshifts.

\begin{table}[tb]
  \begin{center}    
    \begin{tabular}{c c c c c} \hline
      collapse redshift
& 0 & 2 & 5 & 9 \\ \hline \hline
      $\delta_{\rm c}$ assuming PMF model& 1.78 & 1.93 & 2.22 & 2.70 \\
     $\delta_{\rm c}$ assuming $\Lambda$CDM model & 1.61 & 1.59 & 1.58 & 1.58 \\ \hline
    \end{tabular}   
    \caption{The critical over-densities. \:}
  \end{center}
   \label{tab.delc}
\end{table}

\section{Summary and Discussion}
We studied the effect of primordial magnetic fields on the spherical gravitational collapse.
We introduced the angle averaged Lorentz force in the spherical collapse model,
and calculated the non-linear evolution of density perturbations generated by primordial magnetic fields.

The density perturbations of baryon evolve faster than those of dark matter due to the Lorentz force,
and collapses when dark matter density perturbations are still growing.
Therefore, after the virialization of the baryon, we modeled the baryon over-dense region to change its size to maintain the virial equilibrium.
When the baryon over-dense region collapses at a low redshift, dark matter is in the contracting phase and falls into the baryon over-dense region.
Thus, the gravitational potential is enhanced in the virialized baryon region and the baryon over-dense region contracts. 
However, this effect is very small, because the change of gravitational potential is small compared  to the gravitational potential 
which already existed in the baryon system before the dark matter infall after the baryon collapse.
When the baryon over-dense region reaches the virial equilibrium at a high redshift, on the other hand, the dark matter over-dense region is still expanding and the dark matter component gets out of the baryon over-dense region. The baryon system hence loses the gravitational potential support and results in the expansion. 

We compared the virial radius, which we defined to be the radius at the moment of collapse in existence of PMFs,
and the half of the radius at turnaround, which defines the conventional virial radius, for the dark matter over-dense region.
We consider that the virialization for the dark matter over-dense region is reached 
when its radius catches up with that of the baryon over-dense region.
The baryon over-dense region stays smaller the dark matter over-dense region due to PMFs. 
The virial radius for the dark matter over-dense region therefore is always smaller than $R_{\rm max}/2$, 
and becomes even smaller at a higher redshift with the bigger influence of PMFs.

We calculated the critical over-density and the virial over-density of the density perturbations produced by primordial magnetic fields.
The critical over-density reaches $\delta_{\rm c}\simeq1.78$ at the collapse time $a_{\rm coll}=1$, 
and this becomes larger at an earlier collapse time in PMF model.
This is because the effect of magnetic fields is more significant in the early linear evolution than in the non-linear evolution stage.
In contrast, its value is around  $\delta_{\rm c}\simeq1.61$ at $a_{\rm coll}=1$ in the $\Lambda$CDM model.
In the fully matter dominated cosmological model, this is given as $\delta_{\rm c}\simeq1.69$.
We define $\delta_{\rm c}$ at the time of virialization rather than at the time when the density perturbations goes to infinity,
therefore our estimation becomes smaller than the conventional value, $\delta_{\rm c}\simeq1.69$.
The virial over-density $\Delta_{\rm vir}$ in the PMF model becomes larger than that in the $\Lambda$CDM model, and it is about 5 times larger at $a_{\rm coll}=1.0$ and 60 times larger at $a_{\rm coll}=0.1$.
Thus, the structures formed in the PMF model are denser and more compact than those in the $\Lambda$CDM model.
The behaviors of $\Delta_{\rm vir}$ is analogous in both models at a low redshit because $\Delta_{\rm vir}$ is mainly determined by the gravity of dark matter. On the other hand, since the effect of magnetic fields becomes more significant for the earlier collapse,
$\Delta_{\rm vir}$ becomes larger at a higher redshift.

The critical over-density is the threshold value for non-linear gravitational collapses, and characterizes the dark matter halo mass function.
We calculated the PS mass function of the power spectrum of the density perturbations generated by primordial magnetic fields.
The difference in $\delta_{\rm c}$ between the PMF model and the $\Lambda$CDM model is small at low redshifts, 
and therefore the effects of the PMFs show up only at the high mass tail of the mass function.
At high redshifts, on the other hand, the difference in $\delta_{\rm c}$ becomes large enough to suppress the abundance of halos 
over a wide range of mass scales. The suppression becomes as large as $\sim10^5$ at $z=9$.
Thus, simply using the canonical critical over-density $\delta_{\rm c}\simeq1.69$ in the mass function 
would lead to the significant overestimates of the halo abundance  at a high redshift.

The effect of magnetic fields in the non-linear evolution of the density perturbations hence cannot be ignored 
for the studies of the structure formation in existence of primordial magnetic fields.

\begin{acknowledgments}
We thank P. Coles and M. Oguri for the useful discussions and suggestions. 
This work has been supported in part by Grant-in-Aid for
Scientific Research No. 24340048 (KI) from the Ministry of Education,
Sports, Science and Technology (MEXT) of Japan. 
\end{acknowledgments}

\bibliography{reference}
\end{document}